\documentclass[letterpaper, 10 pt, conference]{ieeeconf}  

\IEEEoverridecommandlockouts                              

\usepackage{graphics} 
\usepackage{epsfig} 
\usepackage{color}

\title{\LARGE \bf
Patch-Based Cervical Cancer Segmentation\\ using Distance from Boundary of Tissue
}

\author{Kengo Araki$^{1}$, Mariyo Rokutan-Kurata$^{2}$, Kazuhiro Terada$^{2}$, Akihiko Yoshizawa$^{2}$, and Ryoma Bise$^{1}$
\thanks{$^{1}$ with Kyushu University,
        {\tt\small bise@ait.kyushu-u.ac.jp, kengo.araki@human.ait.kyushu-u.ac.jp}}%
\thanks{$^{2}$ with Kyoto University Hospital}%
}

\begin{document}

\maketitle
\thispagestyle{empty}
\pagestyle{empty}

\begin{abstract}
Pathological diagnosis is used for examining cancer in detail, and its automation is in demand. To automatically segment each cancer area, a patch-based approach is usually used since a Whole Slide Image (WSI) is huge. However, this approach loses the global information needed to distinguish between classes. In this paper, we utilized the Distance from the Boundary of tissue (DfB), which is global information that can be extracted from the original image. We experimentally applied our method to the three-class classification of cervical cancer, and found that it improved the total performance compared with the conventional method.
\end{abstract}

\section{INTRODUCTION}
Pathological diagnosis is used to precisely examine cancer, and the demand for its automation and a Computer Aided Diagnosis (CAD) system has increased in recent years. To meet this demand, many methods have been developed for segmenting the tumor regions or classifying the types of cancer. In such methods, a Whole Slide Image (WSI), which is a digital slide image captured by a scanner, is generally used for pathological diagnosis.

WSI is a gigapixel image that contains the magnified images of the tissue. It enables us to observe precise features of tissues such as the shape or texture of an individual cell. 
In actual diagnosis, pathologists observe cellular features and staining degree at high magnification. On the other hand, they observe the distribution of features and overall appearance at low magnification. Through such multifaceted observations at various magnifications, they are able to diagnose the class and stage of cancer and identify the factors of their decisions.

WSI is too large to be inputted into a Convolutional Neural Network (CNN), which has been widely used for many pattern recognition tasks including pathological image segmentation. To overcome this constraint, WSI is usually cropped into patch images. Each patch is inputted into the model, and then the class is predicted individually. However, this cropping process will lose the global information, such as the cropped location in the tissue and its adjacent features. The loss of this information may lead to a decrease in the performance of the classification model. 

Particularly, cervical cancer, which is the target of this paper, has different distribution trends in each class; the stage increases as malignancy develops in the epithelium (which is located at the boundary of tissue in WSI), and the cancer stage is reached when the malignancy invades the inside of the tissue. To be more specific, the early stage of cervical cancer only exists around the boundary and rarely exists inside. This distributional characteristic indicates that the location of the patch gives powerful prior information for classifying the cervical cancer stage.

\begin{figure*}[t]
    \begin{center}
        \includegraphics [keepaspectratio, width=0.95\linewidth]{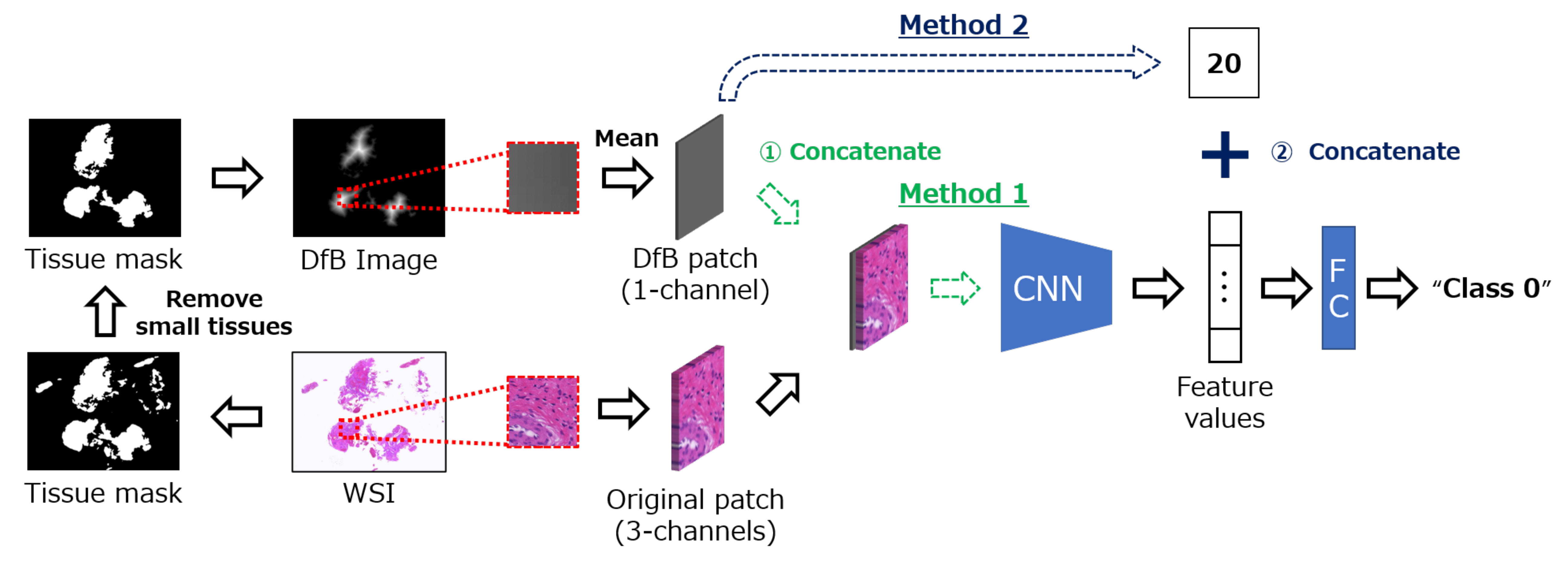}
    \vspace{-3mm}
        \caption{Overview of proposed method. DfB patch is cropped from the same region of Original patch. All pixels of DfB patch are converted into its averaged value. Method1 concatenates the averaged DfB patch with Original patch before inputting it into CNN. Method2 concatenates the single averaged value with feature values outputted from CNN. After concatenation, these feature values are inputted into Fully Connected (FC) layers to classify.}
        \label{fig:propsed_method}
    \end{center}
    \vspace{-3mm}
\end{figure*}

In this paper, we focus on the differences in the distribution characteristics between the stages. To quantify this characteristic, we used the Distance from the Boundary of tissue (DfB). A DfB is a value for each pixel that indicate the distance from the boundary of the tissue; a distant pixel from the boundary ({\it i.e.}, a pixel located near the center of the tissue) has a higher value as shown in Fig. \ref{fig:propsed_method}. In other words, the pixel value of DfB provides the information of the pixel location in the tissue.
In our method, we simply introduce this DfB value into the patch-level classification task for pathological image segmentation.


Our main contributions in this paper are as follows: (1) To deal with the problem of the global information lost by cropping it into patches, we introduce the DfB value into the patch classification task in pathological image segmentation, in which DfB contains the distribution characteristics that differ among classes. (2) We found that the effectiveness of the distance information in cervical cancer classification varies with the distance from the boundary. (3) Our proposed method improves the prediction performance for the Non-Neoplasm class.

\begin{figure}[t]
    \begin{center}
        \includegraphics [keepaspectratio, width=0.93\linewidth]{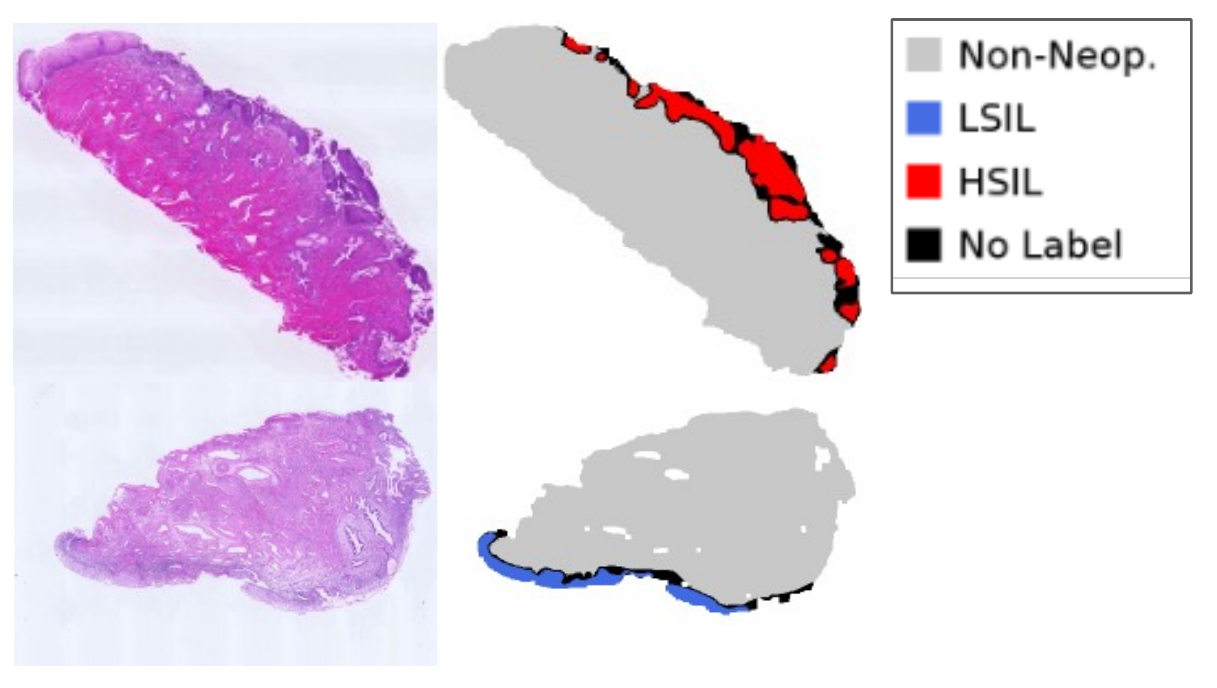}
        \caption{Examples of Whole Slide Image (WSI) and ground truth for cervical cancer. We can observe that the tumor regions (LSIL and HSIL) appear around the boundary of the tissue.}
        \label{fig:example_gt}
    \end{center}
    \vspace{-3mm}
\end{figure}

\begin{figure}[th]
    \begin{center}
        \includegraphics [keepaspectratio, width=0.9\linewidth]{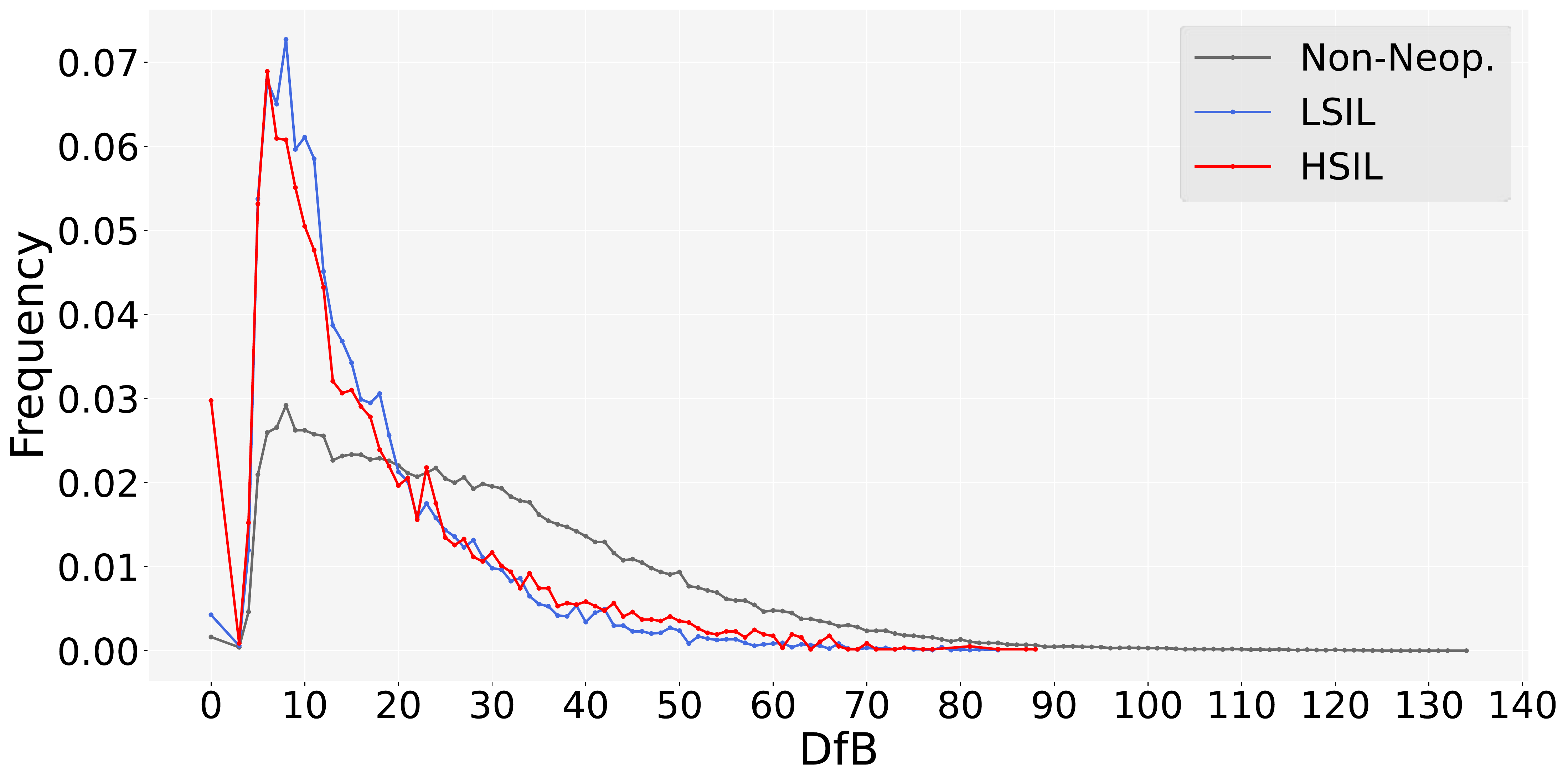}
    \vspace{-3mm}
        \caption{Patch ratio per DfB in each class. The DfB ranges of Non-Neop., LSIL, and HSIL class data are 0 - 133, 0 - 83, and 0 - 88 respectively. LSIL and HSIL incline to locate at the lower DfB area (where is around the boundary).}
        \label{fig:patch_ratio}
    \end{center}
    \vspace{-3mm}
\end{figure}

\section{RELATED WORKS}
Many methods have been proposed for pathological image analysis, where most of them take a patch-based approach. The patch-based methods segment the huge image (WSI) into small patches and then classify each patch image~\cite{AltunbayD2010}\cite{ChangH2014}\cite{CruzRoaA2014}\cite{MousaviHS2015}. 
However, patch-based approaches have a trade-off: the patch with high magnification has detailed features like the shape and texture of cells, while it loses the features of the surrounding area. To handle this problem, using multi-scale features enables the model to capture both detailed features and wide-ranging texture patterns, adaptively \cite{tokunaga2019adaptive}. Even if this \cite{tokunaga2019adaptive} approach uses a wide field of view with low resolution, the location information in the entire image it is still difficult to use since a WSI is significantly large.

A distance map is used for individual cell segmentation tasks in pathological images.
Cells in pathological images are densely distributed and their boundary is blurry, which may lead to segmenting some cells as one instance. 
To segment individual cells, Neylor et al. \cite{naylor2018segmentation} solved this segmentation task by multi-task learning that simultaneously predicts segmentation of cells and regression of the intra-nuclear distance maps. Multi-task learning facilitates the method to identify the boundary of individual cells. Their purpose for using a distance map is different from ours: we use distance information of an entire tissue.




\section{DISTANCE FROM BOUNDARY OF TISSUE}

In this study, we classify each patch image into three classes; 1. Non-Neoplasm (Non-Neop.) 2. Low Squamous Intraepithelial Lesion (LSIL), 3. High Squamous Intraepithelial Lesion (HSIL).
We investigated the prior of the DfB value for each class.
Fig. \ref{fig:patch_ratio} shows the distribution of the DfB value for each class, in which the horizontal and vertical axes indicate the DfB value and the normalized frequency of patches in each distance, respectively.
In this graph, we can observe that LSIL and HSIL are biased toward the boundary of the tissue, whereas Non-Neop. is widely distributed over the whole area. Naturally, the frequency of smaller distances tends to be high, since the patches near the boundary have a larger circumference than the patches inside.
We effectively use this information for classifying patch images.

\section{PROPOSED METHOD}

Fig. \ref{fig:propsed_method} shows an overview of the proposed method that utilizes the DfB as prior information. 
To segment the huge image (WSI), we follow a patch-based classification approach.
To address the problem of losing the cropped position of the patches, we use the distance of the patch from the tissue's boundary as input since DfB gives powerful prior information for the classification as discussed above.
To input DfB into the model, we proposed two methods. One concatenates a DfB patch with the original patch before inputting it into the CNN model. The other concatenates the mean of DfB value with feature values before inputting it into Fully Connected (FC) layers.

\subsection{Distance transform}
To make a DfB image, we first roughly segment the tissue region by thresholding. In this process, we make a scaled-down WSI (from magnification-40x to magnification-2.5x)  since the original magnification of WSI is too large to process image transformation and the detailed distance information is not so important.
Next, we set the thresholds in Hue, Saturation, Value (HSV) color space. The tissue area and background area are segmented into a binary image as shown in Fig. \ref{fig:propsed_method}.
Then, small objects are removed to get rid of fragments of tissue and fill the small holes to avoid a negative effect on the distance representation. 

After obtaining the tissue mask, we make the DfB image by using distance transformation \cite{borgefors1986distance} from the tissue mask.
The values of all the pixels in the background area are 0, and the value of the pixel inside of the tissue is the distance from the boundary of the foreground region in the tissue mask, in which the unit of the distance is the pixel distance in the resized image. In our dataset, the maximum value was 140.

\subsection{Network}
Our method has two ways to input DfB into the model: one is inputting a DfB patch into a CNN with an original patch (Method1 in Fig. \ref{fig:propsed_method}), and the other is inputting a DfB value into FC layers with feature values that are extracted from the CNN (Method2 in Fig. \ref{fig:propsed_method}). Before inputting DfB, a DfB patch is cropped from the position corresponding to its original patch. Here, the location where the patch is cropped is important information, while the detailed location in a patch is not necessary. Therefore, the DfB patch converts into the mean of its values. This model outputs the probabilistic score of each class through a Softmax function after FC layers.
For the loss function of the CNN and FC layers, we used a categorical cross-entropy for the output of the Softmax function. 

In the inference process, we make a prediction map to visually understand which area is well classified. A prediction map is made as follows: First, WSI is cropped into patches with a stride equal to the patch size, and the corresponding DfB is concatenated in the same manner as the training step. The CNN predicts the class of each patch.
Finally, prediction classes of each patch are merged and masked the area that is not subject to classification or background.

\section{EXPERIMENTS}
We evaluated our method on three-class (Non-Neop., LSIL, and HSIL) classification for WSIs of uterine cervix. For this experiment, we compared the confusion matrices and five metrics with four methods: Baseline, DfB + CNN, DfB + FC, and DfB + FC (Transfer). For the initial weights of DfB + FC (Transfer), we used the weights of the pre-trained baseline's model and updated all the weights using DfB. 

\subsection{Experimental setup}
Images of a sliced uterine cervix stained by Hematoxylin and Eosin (H\&E) were captured by a virtual slide scanner with a maximum magnification of 40x. In the experiment, we used 282 WSIs. To generate the ground truth of the dataset, three pathologists manually annotated the regions of three cervical cancer grades: 1. Non-Neop., 2. LSIL, and 3. HSIL. 

To train our model, WSI was cropped into patches with 256 $\times$ 256 pixels window size, 256 pixels stride size, and 40x magnification. The patches were randomly flipped along the horizontal and vertical axes for data augmentation. To handle class imbalances, we randomly remove samples from the majority class (under-sampling) and add more samples by duplicating the existing one from the minority class (over-sampling). This data augmentation process was only executed for the training set.

We experimented using five-fold cross-validation; 282 WSIs (194,425 patch images) were divided into 5 sets (56 or 57 WSIs per fold). One set was used in the test, and the other sets were used for training. 20\% of WSIs in the training set were used as validation data, and the rest was used as training data. Validation data was used as a criterion for selecting the best performed model. After that, each WSI was split into patches. Only patches that consisted of a single class were used for training and testing except for the process of making prediction maps.

We used ResNet-50 \cite{he2015deep} for CNN and Adam \cite{kingma2017adam} for the optimizer with a learning-rate of $10^{-3}$. ResNet-50 was pretrained on ImageNet \cite{deng2009imagenet}. The optimization was suspended when the mRecall did not improve in five epochs.

We evaluated five metrics: Accuracy, mean of the per-class Recall (mRecall), mean of the per-class Precision (mPrecision), F1-score (F1), and mean of the Intersection of Union (mIoU). They are defined as: Accuracy = $\frac{\sum_{c}TP_c}{\sum_{c}(TP_c + FN_c)}$,
mRecall = $\frac{1}{M}\sum_{c}\frac{TP_c}{TP_c + FN_c}$, mPrecision = $\frac{1}{M}\sum_{c}\frac{TP_c}{TP_c + FP_c}$, 
F1 = $2 \times \frac{mRecall \times mPrecision}{mRecall + mPrecision}$, mIoU = $\frac{1}{M}\sum_{c}\frac{TP_c}{TP_c + FP_c + FN_c}$,
where $M$ is the number of classes, and $TP_c$, $FP_c$, and $FN_c$ are the numbers of true positives, false positives, and false negatives for class $c$, respectively.

\begin{figure*}[tbhp]
    \begin{center}
        \includegraphics [keepaspectratio, width=0.9\linewidth]{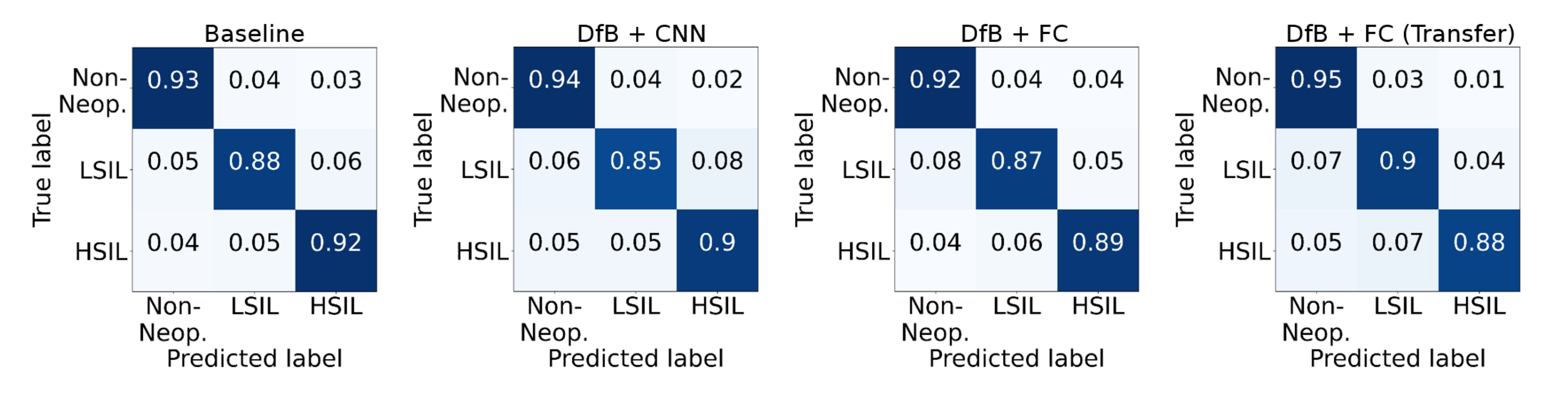}
    \vspace{-2mm}
        \caption{Comparison of confusion matrices for three-class classification. Diagonal components of a matrix is a recall for each class. From left to right, Baseline, DfB+CNN, DfB+FC, and DfB+FC (Transfer).}
        \label{fig:confusion_matrix}
    \end{center}
    \vspace{-3mm}
\end{figure*}

\subsection{Experimental Results}

Fig. \ref{fig:confusion_matrix} shows the confusion matrices of three-class classification. These are the aggregated results for each test set of cross-validation. Each element of the matrix is normalized by the total number of its true labels, and the diagonal components represent recall of each class. In DfB + CNN and DfB + FC (Transfer), the score of Non-Neop.-class was better than the baseline's score. Moreover, in DfB + FC (Transfer) improved LSIL score.

Table \ref{table:compare_metrics} shows the performance of these metrics for each ablation method. Our model DfB + FC (Transfer) achieved the highest scores in all metrics. In particular, mPrecision, and F1 were improved by 7.3\% and 4.6\%, respectively. 
We consider that the prior about the location information of each patch facilitates improving the performance.
In contrast, DfB + FC deteriorated in all scores. It is assumed to be difficult to simultaneously learn the combination of image features and DfB value without pretraining.

\begin{table}[t]
    \begin{center}
        \caption{Comparison of four methods in three-class (Non-Neop., LSIL, and HSIL) classification.}
        \begin{tabular}{lccccc}
        \hline
        Method & Acc. & mRecall & mPrec. & F1 & mIoU \\ \hline \hline
        Baseline    & 0.926   & 0.909   & 0.678   & 0.777   & 0.834   \\
        DfB + CNN   & 0.935   & 0.898   & 0.697   & 0.785   & 0.814   \\
        DfB + FC    & 0.917   & 0.896   & 0.658   & 0.758   & 0.811   \\
        \multicolumn{1}{c}{DfB + FC (Transfer)}   & \textbf{0.948}   & \textbf{0.911}   & \textbf{0.751}   & \textbf{0.823}   & \textbf{0.836}   \\ \hline
        \end{tabular}
        \label{table:compare_metrics}
    \end{center}
\end{table}

\begin{figure}[t]
    \begin{center}
        \includegraphics [keepaspectratio, width=0.9\linewidth]{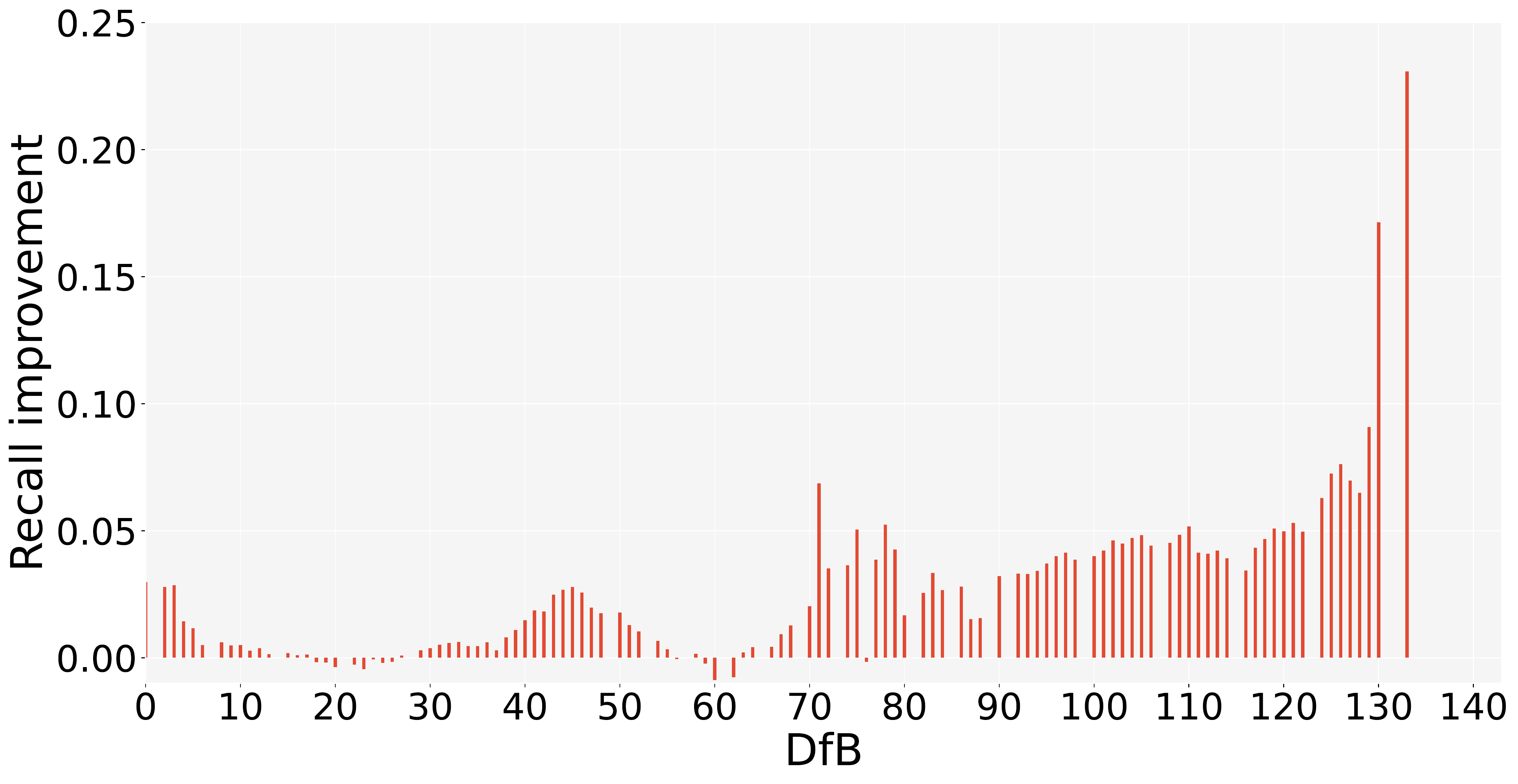}
    \vspace{0mm}
        \caption{Difference in mean recall between DfB+FC (Transfer) and Baseline in each distance. Higher than 0.00 in y-axis indicates that DfB+FC (Transfer) is better than Baseline.}
        \label{fig:acc_diff_all}
    \vspace{0mm}
    \end{center}
\end{figure}

Fig. \ref{fig:acc_diff_all} shows the improvement of mean recall by using DfB+FC (Transfer) in each value of DfB. The horizontal axis indicates the DfB value, and the vertical indicates the improvement over the baseline at a certain distance.
If the value of the y-axis is greater than 0, our method improved on the DfB. 
We can observe that our method improved on the baseline at almost all distances, in particularly at the larger values.
Since there are no patches with DfB greater than approximately 90 in LSIL and HSIL, the improvement of our method at the larger values was significant.

\begin{figure}[t]
    \begin{center}
        \includegraphics [keepaspectratio, width=\linewidth]{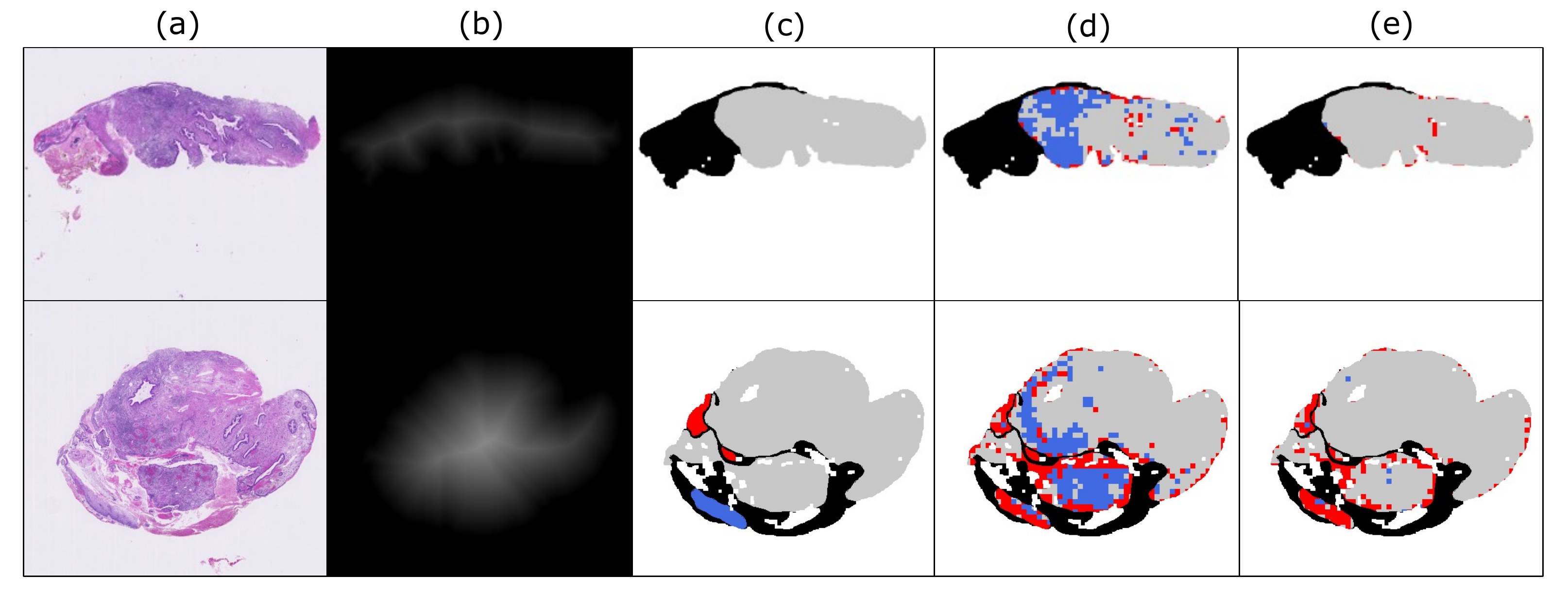}
        \caption{Examples of prediction results. (a) Original image, (b) DfB image, (c) Ground truth, (d) Baseline, and (e) DfB+FC (Transfer). Gray, blue, red, and black for (c), (d), (e) indicates Non-Neop., LSIL, HSIL, and No-label area, respectively.}
        \label{fig:predmap}
    \end{center}
    \vspace{-4mm}
\end{figure}

Fig. \ref{fig:predmap} shows the examples of prediction results. This is the comparison of DfB + FC (Transfer) and Baseline. In (b) DfB image, the brighter area has larger DfB values, which means the pixel is far from the boundary. As shown in Fig. \ref{fig:acc_diff_all}, the number of misclassified patches of Non-Neop. decreased. The reason for misclassification as LSIL in (d) Baseline is considered to be that a similar image feature's patch labeled as LSIL is included in the training set. The prior information, ({\it i.e.}, LSIL patches rarely exist in the interior of the tissue) works for predicting correctly.

\section{CONCLUSIONS}
We proposed a segmentation method in pathological images (WSI) that utilizes the Distance from the Boundary of the tissue (DfB). The proposed method can address the problem of the patch-based methods that lose the cropped position of the patches by using the distance of the patch from the tissue’s boundary as input with the original patch. The experiments demonstrated the effectiveness of our method; it improved the F1 by 4.6\% compared with the baseline method. In particular, the classification performance of Non-Neop. and LSIL, which are difficult even for pathologists, had improved. The proposed method might be applied to other squamous cell carcinomas as well.


\vspace{8mm}
\noindent {\bf Acknowledgments:}
This work was supported by AMED under Grant Number JP20lk1010036h0002 and JSPS KAKENHI Grant Numbers JP20H04211.

\bibliography{ref}
\bibliographystyle{unsrt}

\end{document}